# Updatable Learned Index with Precise Positions

Jiacheng Wu
Tsinghua University
Beijing, China
wu-jc18@mails.tsinghua.edu.cn

Yong Zhang*
Tsinghua University
Beijing, China
zhangyong05@tsinghua.edu.cn

Shimin Chen*
Chinese Academy of Sciences
Beijing, China
chensm@ict.ac.cn

Jin Wang
UCLA
Los Angeles, USA
jinwang@cs.ucla.edu

Yu Chen
Tsinghua University
Beijing, China
y-c19@mails.tsinghua.edu.cn

Chunxiao Xing
Tsinghua University
Beijing, China
xingcx@tsinghua.edu.cn

## ABSTRACT

Index plays an essential role in modern database engines to accelerate the query processing. The new paradigm of "learned index" has significantly changed the way of designing index structures in DBMS. The key insight is that indexes could be regarded as learned models that predict the position of a lookup key in the dataset. While such studies show promising results in both lookup time and index size, they cannot efficiently support update operations. Although recent studies have proposed some preliminary approaches to support update, they are at the cost of scarifying the lookup performance as they suffer from the overheads brought by imprecise predictions in the leaf nodes.

In this paper, we propose LIPP, a brand new framework of learned index to address such issues. Similar with state-of-the-art learned index structures, LIPP is able to support all kinds of index operations, namely lookup query, range query, insert, delete, update and bulkload. Meanwhile, we overcome the limitations of previous studies by properly extending the tree structure when dealing with update operations so as to eliminate the deviation of location predicted by the models in the leaf nodes. Moreover, we further propose a dynamic adjustment strategy to ensure that the height of the tree index is tightly bounded and provide comprehensive theoretical analysis to illustrate it. We conduct an extensive set of experiments on several real-life and synthetic datasets. The results demonstrate that our method consistently outperforms state-of-the-art solutions, achieving by up to 4×for a broader class of workloads with different index operations.

## 1 INTRODUCTION

Tree indexes are essential components to support efficient data access in modern database engines. Many different index structures have been proposed to meet the requirement of various access patterns and workloads. The recent study on *Learned Index* [22] has opened up a new way to construct the index for sorted data. Given a dataset, Learned Index utilizes machine learning models to learn the data distribution and predict the position of a lookup key in the dataset. It could be realized via supervised learning techniques by using the Cumulative Distribution Function (CDF) of the dataset for training. Since the models might be inaccurate for the predicted positions, the learned index needs to search the lookup key in a bounded range around the predicted positions. Recent comprehensive experimental studies [20, 31] demonstrate that learned indexes achieve significant advantages over traditional index structures in terms of high performance and low memory footprint.

Nevertheless, the original Learned Index [22] only supports lookup on read-only datasets and fails to handle update operations which are essential in index structures. To address this problem, two recent studies namely ALEX [11] and PGM [12] propose several strategies to add support for updating the index. However, their support for updates is at the expense of extra search overhead for lookup operations. PGM uses the logarithmic method and thus needs to find keys in a series of subtrees instead of a single one. What's more, ALEX even has unbounded "last mile" search cost in the leaf nodes since it does not provide any threshold of errors caused by the wrong predictions. As illustrated in the example shown in Figure 1, the lookup time of ALEX is dominated by the search in the leaf nodes, which in the worst case would need linear or binary search on the whole node. Consequently, they may suffer from the poor lookup performance. Moreover, the update operations on these indexes also incur huge amounts of elements shifting. These overheads are all brought by the imprecise predictions of learned models.

To tackle with these issues brought by inaccurate predictions, we propose the Updatable **L**earned **I**ndex with **P**recise **P**ositions (LIPP), a brand new learned index to provide efficient support for a full set of index operations, namely lookup query, range query, insert, update, delete and bulkload. A distinct advantage of LIPP is that it eliminates the "last mile" search in the leaf nodes, thereby bounding the lookup cost to tree height and significantly improving index performance. The key-to-position mapping is precise in LIPP. If multiple keys are mapped into the same position, a new child node will be created to hold the keys. To bound the height of the tree index, we propose kernelized linear models that are able to evenly distribute the mapping of newly inserted elements to positions, and a light-weight adjustment strategy to keep the tree height bounded. We also provide theoretical guarantee that the height of LIPP is bounded to $O(\log N)$ with $N$ as the cardinality of dataset.

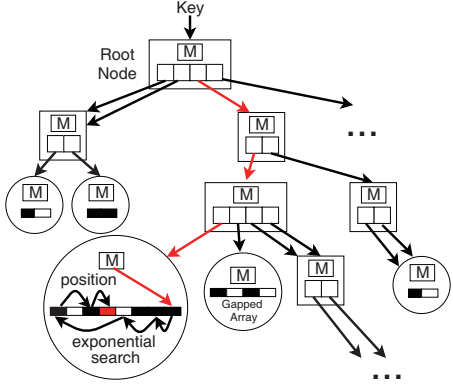

Figure 1: Extreme Case for ALEX [11]

We conduct an extensive set of experiments on both real world and synthetic datasets with workloads of mixed operations. The experimental results show that LIPP outperforms state-of-the-art learned index structures by an obvious margin. Specifically, LIPP achieves 2.8× and 6.3× better lookup performance than ALEX and PGM for read-only workload. In terms of write-only and read-write workloads, LIPP outperforms ALEX by up to 2.9×, and has similar index size under most settings.

The rest of this paper is organized as follows: Section 2 introduces the background for learned index series. Section 3 describes the structure and properties of LIPP. Section 4 displays the details of operations on LIPP. Section 5 provides theoretical analysis on the performance of LIPP. Section 6 presents the experimental results. Section 7 discusses issues related to concurrency and new hardware accelerators. Section 8 reviews the related work. Finally, the conclusion is made in Section 9.

## 2 PRELIMINARIES
### 2.1 Tree-Based Index
In this paper, we aim at devising a learned index structure that can support all operations in traditional tree-based indexes. Given $x$ and $y$ as keys and $v$ as value, an index $S$ supports the operations:
(1) $member(x)$ = TRUE if $x \in S$, FALSE otherwise.
(2) $lookup(x)$ returns the element with key $x \in S$ (if any), NIL otherwise.
(3) $range(x, y)$ returns the elements whose keys $\in [x, y]$
(4) $insert(x, v)$ inserts the element with key $x$ and value $v$ to $S$.
(5) $delete(x)$ removes the element with key $x$ from $S$.
(6) $update(x, nv)$ is implemented with the $delete(x)$ followed by $insert(x, nv)$ with a new key or value.
(7) $bulkload(x[N])$ is used in practice to index $N$ elements at initialization or for rebuilding index.

In this paper, we assume that the keys are unique. It is easy for indexes to support duplicate keys, e.g. maintaining a pointer to an overflow list. A typical example of tree-based structure is B+Tree, a dynamic height-balanced tree. The overhead for lookup operations on B+Tree consists of traversing from root to leaf nodes and binary search in nodes. Besides, the node size of B+Tree is also limited. The large node size will result in the huge cost of searching keys inside nodes.

### 2.2 Learned Index
Given a key, Learned Index [22] maps it to the position in the sorted array of keys, which thus is considered as a trained model. Learned Index builds a hierarchy of models with fixed height called Recursive Model Index (RMI). In order to locate a key, the higher-level model predicts the model at the next level by learning the CDF of the dataset. And the leaf-level model outputs the final prediction for the position of the key. Finally, Learned Index applies extra binary search to correct the wrong predictions on the sorted array based on the given error bound $\epsilon$. Due to the error bound, Learned Index has much larger nodes that hold many more elements but without incurring drastically higher search cost compared with B+Tree. The larger node size of Learned Index results in fewer levels of index, which significantly saves the cost of index traversal. This is the main reason why Learned Index outperforms B+Tree in lookup operations. However, Learned Index cannot support updates.

Some recent studies aim at supporting updates. PGM [37] uses linear models and separates the keys in different linear segments with given error bound. Unlike Learned Index, each model or segment of PGM specifies the first key covered by the segment. In this way, PGM recursively constructs index on the sorted keys of segments in low level. To locate a key, PGM applies the similar traversal process as Learned Index except that the predicted position at each level is required to be corrected immediately. To support insertions, PGM employs the idea of LSM-tree [38]. Concretely, keys are separated into subsets with different sizes and PGM indexes are built over those subsets. Each insertion of a key is required to find a series of non-empty sets, merge them into a large subset and build a new PGM index on the large one. Unfortunately, it decreases the lookup performance heavily since it has to search for a given key in all components with different sizes. As shown in Table 1, the lookup operations of PGM cost $O(\log^2 N)$.

Meanwhile, ALEX separates elements in many data nodes shown in Figure 1. The lookup procedure of ALEX is similar to Learned Index, i.e., recursively locating the models in the next level. However, ALEX uses exponential search in the leaf node to locate the given key, which is due to that ALEX is not bounded by any prediction error threshold. The insert procedure of ALEX first utilizes the lookup procedure to locate a proper position for inserted key in the nodes and then tries to the insert the key into that position. ALEX also shifts the elements to make the gap for the inserted key when there is a key in that position. As shown in Table 1, the shifting procedure costs $O(\log m)$ on average due to the underlying layout gapped array of each data node, but can have $O(m)$ cost in the worst case, where $m$ is the max node size. Furthermore, ALEX performs node expansion or split when a leaf node does not have enough free space for an insertion, but fails to provide upper bound of time complexity on the performance for insertions. As shown in Table 1, our solution LIPP avoids such problems and thus has lower complexity and average latency.

Table 1: Summary of Complexity and Average Latency Comparisons among Different Indexes

| | | LIPP | ALEX | PGM | Learned | B+Tree |
|---|---|---|---|---|---|---|
| Lookup | Complexity | $O(\log N)$ | $O(\log N + \log m)$ [1] | $O(\log^2 N)$ | $O(\log N)$ | $O(\log N)$ |
| | Latency [6] | 24.23ns | 68.92ns | 151.53ns | 139.09ns | 237.94ns |
| Insert | Complexity | $O(\log^2 N)$ | $O(\log^2 N + \log m)$ [2] | $O(\log^2 N + \log N)$ [3] | — | $O(\log N)$ |
| | Latency [6] | 70.93ns | 204.94ns | 217.17ns | — | 1114.19ns |
| Search Range (Leaf) | | $O(1)$ [4] | $O(m)$ | $O(\epsilon)$ [5] | $O(\epsilon)$ | $O(m)$ |
| Search Range (Non-Leaf) | | $O(1)$ | $O(1)$ | $O(\epsilon)$ | $O(1)$ | $O(m)$ |

[1] $m$ is the max number of slots in nodes. [2] $O(\log^2 N) + O(m)$ for extreme cases. [3] Need existence checking to prevent duplicated keys. [4] $O(1)$ means no need to search in nodes. [5] $\epsilon$ is the prediction error threshold. [6] The latency is conducted on YCSB.

### 2.3 Monotonically Increasing Models

A *Monotonically Increasing* function is one defined on ordered sets that preserves the given order. Given a monotonically increasing model $\mathcal{M}$, for any pair of keys $k_i$ and $k_j$, the property in Equation (1) holds:

$$k_i \le k_j \rightarrow \mathcal{M}(k_i) \le \mathcal{M}(k_j) \tag{1}$$

It is essential for the learned index structures to satisfy this property so as to support range queries. A range query first locates the position of the start key, then scans forward until it reaches the end key. However, if the model is not monotonically increasing, it may map the start key to a position after the predicted position of the end key, causing incorrect results for the range query. In this paper, we follow previous studies [11, 22] to employ linear models for prediction, which satisfy the monotonically increasing property.

In our approach, the learned model $\mathcal{M}$ is a *kernelized linear function*. We need to store the kernel function $\mathcal{G}$ and two model parameters, i.e., the slope $A$ and the intercept $b$. Given a key $k$, the model computes the entry position in the node with an array of $L$ entries as Equation (2):

$$\mathcal{M}(k) = \begin{cases} 0 & \lfloor A \cdot \mathcal{G}(k) + b \rfloor < 0 \\ L - 1 & \lfloor A \cdot \mathcal{G}(k) + b \rfloor \ge L \\ \lfloor A \cdot \mathcal{G}(k) + b \rfloor & otherwise \end{cases} \tag{2}$$

The only requirement of the kernel function is that it must be monotonically increasing. Examples of kernel functions $\mathcal{G}$ include the exponential function $e^x$, the logarithm function $\ln(x)$, the linear function $x$, the quadratic function $x^2$ when keys are positive, and even polynomial functions $\sum c_i x^i$, etc. In many real-world applications, we find the linear function behaves well. Accordingly, we use the linear function as the default kernel function, unless otherwise stated. However, if the distribution of a target data set is not regular, our approach allows users to take advantage of the prior knowledge of dataset and to specify a kernel function to improve the performance of model. Due to the space limitation, we leave the experiments related to the effect of specified kernel functions of our methods on synthetic dataset in Appendix [44] B.1.

## 3 THE LIPP INDEX

### 3.1 Overview

The core idea of LIPP is to avoid inaccurate predictions, i.e., **all predictions made by models are exact.** With precise positions, the significant and inevitable overheads, including in-node search for lookup, element shifting for insertion, can be eliminated. Nevertheless, to reach this goal we need to overcome the following two challenges:

Firstly, **predictions for two different keys might coincide at one position**. We call elements with such keys as *conflicting* elements. To resolve conflicts, ALEX shifts the elements to make a gap for the newly inserted ones, which causes inaccurate predictions. Instead, we preserve the precise predictions by replacing the current element with a new node that accommodates these two conflicting elements (cf. Section 4.1 and 4.2).

The second challenge is derived from the first one: **Simply creating new nodes for conflicting keys would cause the tree height increasing without bound**, thus hurting the performance for both index lookup and insert operations. To deal with this challenge, we propose a novel adjustment strategy which redistributes keys in a subtree to control the height of the subtree. It can wisely select the appropriate subtree and determine when and how to adjust the subtree to reduce the tree height (cf. Section 3.3 and 4.3).

We also provide theoretical guarantee for the tree height, along with the complexity of lookup and insert operations (cf. Section 5). As a result, we expect LIPP to be faster than both existing learned index structures and traditional B+Trees for all index operations, while the index size of LIPP is comparable to those of existing learned index structures.

### 3.2 Structure

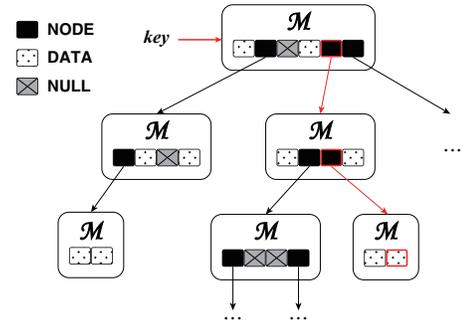

Figure 2: Structure of LIPP

The overall structure of LIPP is shown in Figure 2. Each node contains a model $\mathcal{M}$, an array of entries $\mathcal{E}$, and a bit vector of entry types. There are three types of entries in a node:

(1) **NULL**: The entry is an unused slot (gap). All entries are initialized as *NULL* and used to store new keys. After an insertion to a slot completes, the type of the entry is changed to *DATA*.
(2) **DATA**: The entry contains one element with its key and payload. When the payload is too large, we store a pointer (or an offset) to the payload in the entry.

(3) **NODE**: The entry points to a child node in the next level, thus helping form a tree structures. When a new element is inserted into a *DATA* entry, a child node is created to hold the two conflicting entries. The current entry becomes *NODE* and its content is the pointer to the child node.

The size of all three types of entries is 16 bytes. A *DATA* entry consists of an 8B key and an 8B payload or pointers to payload, while a *NODE* entry contains a child node pointer. The bit vector specifies the type of each entry with two bits. For the $i^{th}$ entry, the $2i^{th}$ bit indicates whether the entry is *NULL* or not, and the $(2i+1)^{th}$ shows the entry type, i.e. *DATA* or *NODE*. Since the bit vector is a light-weighted structure, the node size could be bounded by a predefined hyper-parameter (e.g., 16MB).

Unlike existing learned index structures, LIPP does not distinguish leaf nodes (DataNode) from non-leaf nodes (InnerNode). Instead, all nodes are treated equally. Entry types are used to guide the index operations to choose different actions. Details are illustrated in Section 4.

LIPP is a sorted index. In every node, the entries (either the *DATA* entry or entries in the subtree pointed to by the *NODE* entry) in the array are sorted in key order. That is, keys in the left part of the array are less than keys in the right part whether they are in the node directly or in subtrees. This property is achieved by simply requiring all models in all nodes monotonically increase.

### 3.3 Metrics for Evaluating the Learned Model

In this section, we start with finding a metric to evaluate the quality of a learned model for indexing.

Given a model $\mathcal{M}$, two keys $k_i$ and $k_j$ conflict if and only if $\mathcal{M}(k_i) == \mathcal{M}(k_j)$. We propose the notion of *conflict degree* in Definition 3.1 to capture the maximum number of conflicts at any position in the entry array:

*Definition 3.1.* For a node with $L$ entries and a learned model $\mathcal{M}(k)$, the conflict degree $T_{\mathcal{M}}$ of the node is:
$$T_{\mathcal{M}} = \max_{l \in [0, L-1]} |\{k \in \mathcal{K} | \mathcal{M}(k) == l\}| \quad (3)$$
where $\mathcal{K}$ is the set of keys contained in this node, and $l$ is a possible position ranging from 0 to $L-1$.

According to such a definition, the better the model is, the lower the conflict degree it has. The conflict degree of the ideal model is 1. The worst model maps all keys to the same position with a conflict degree of $|\mathcal{K}|$. Our goal is to find a model to achieve as low conflict degree as possible.

We observe that there exists an upper bound for the minimum $T_{\mathcal{M}}$, i.e. $\exists \mathcal{M}, T_{\mathcal{M}} \leq \lceil \frac{N}{3} \rceil$ where $N$ is the number of keys in $\mathcal{K}$, i.e. $N = |\mathcal{K}|$. However, the $\lceil \frac{N}{3} \rceil$ may not be the tightest upper bound in many cases. Thus, our goal is to find a best model $\mathcal{M} = A\mathcal{G}(k) + b$ with the minimum conflict degree $T_{\mathcal{M}}$. We need to first identify certain properties that the model $\mathcal{M}$ should satisfy under a given conflict degree $T$. The $\mathcal{M}$ needs to map keys to positions between 0 and $L-1$. If a key is mapped to a position beyond the range, we set its position to either 0 or $L-1$. However, since the position 0 or $L-1$ contains at most $T$ elements, the values of parameters $A$ and $b$ should follow Condition (4).
$$\begin{cases} A \cdot \mathcal{G}(k_i) + b \geq 1, & \exists i \leq T \\ A \cdot \mathcal{G}(k_{N-1-j}) + b < L-1, & \exists j \leq T \end{cases} \quad (4)$$

**Algorithm 1:** FMCD($\mathcal{K}$, $L$)

**Input**: $\mathcal{K}$: the collection of keys, $L$: the number of entries
**Output**: $\mathcal{M}$: the model, $T$: the conflict degree

1 **begin**
2     $i = 0; T = 1; N = |\mathcal{K}|$;
3     $U_T = \frac{\mathcal{G}(k_{N-1-T}) - \mathcal{G}(k_T)}{L-2}$;
4     **while** $i \leq N - 1 - T$ **do**
5        **while** $i + T < N$ **and** $\mathcal{G}(k_{i+T}) - \mathcal{G}(k_i) \geq U$ **do**
6           $i = i + 1$;
7        **if** $i + T \geq N$ **then**
8           **break**;
9        $T = T + 1$;
10        $U_T = \frac{\mathcal{G}(k_{N-1-T}) - \mathcal{G}(k_T)}{L-2}$;
11     $\mathcal{M}.A = \frac{1}{U_T}$;
12     $\mathcal{M}.b = \frac{L - (\mathcal{M}.A \cdot (\mathcal{G}(k_{N-1-T}) + \mathcal{G}(k_T)))}{2}$;
13     **return** $\{\mathcal{M}, T\}$;
14 **end**

Since the keys are sorted, the former formula means there exists a key $k_i$ in the first $T$ elements of $\mathcal{K}$ that is not mapped to position 0. In other words, at most $T$ elements go to position 0. The latter formula can be derived in a similar way. Based on the above conditions, we further obtain the constraint on $A$ as Condition (5).
$$A \leq \max_{i,j} \frac{L-2}{\mathcal{G}(k_{N-1-j}) - \mathcal{G}(k_i)} = \frac{L-2}{\mathcal{G}(k_{N-1-T}) - \mathcal{G}(k_T)} \quad (5)$$
Once the value of $A$ satisfies Condition (5), it is easy to find a value $b$ (e.g., $b = 1 - A \cdot \mathcal{G}(k_T)$) to satisfy Condition (4). Therefore, Condition (5) and Condition (4) are essentially equivalent.

Besides, according to the definition of $T$, any consecutive $T+1$ elements should not conflict in the same position. Specifically, whether two keys conflict with each other can be checked by Lemma 3.2.

LEMMA 3.2. $\forall k, k'$, if $A \geq |(\mathcal{G}(k) - \mathcal{G}(k'))|^{-1}$, $k$ and $k'$ will be mapped to different positions.

PROOF. $|A \cdot (\mathcal{G}(k) - \mathcal{G}(k'))| = |(A \cdot \mathcal{G}(k) + b) - (A \cdot \mathcal{G}(k') + b)| \geq 1$. Then $\lfloor A \cdot \mathcal{G}(k) + b \rfloor \neq \lfloor A \cdot \mathcal{G}(k') + b \rfloor$. Thus, $k$ and $k'$ are mapped to different positions. □

Therefore, $A \geq |(\mathcal{G}(k_i) - \mathcal{G}(k_{i+T}))|^{-1}$ ensures $k_i$ and $k_{i+T}$ do not conflict. We can further have the Condition (6).
$$A \geq \max_{i \in [0, N-1-T]} \frac{1}{\mathcal{G}(k_{i+T}) - \mathcal{G}(k_i)} \quad (6)$$
Finally, based on Condition (5) and (6), we conclude that the value of $T$ must follow the Condition (7).
$$\frac{L-2}{\mathcal{G}(k_{N-1-T}) - \mathcal{G}(k_T)} \geq \max_{i \in [0, N-1-T]} \frac{1}{\mathcal{G}(k_{i+T}) - \mathcal{G}(k_i)} \quad (7)$$
Similarly, the value of $b$ should satisfy the Condition (8):
$$L - 1 - A \cdot \mathcal{G}(k_{N-1-T}) \geq b \geq 1 - A \cdot \mathcal{G}(k_T) \quad (8)$$

### 3.4 Efficient Algorithm for Deciding the Model

According to the conclusion in Condition (7), the next step is to design an efficient algorithm to find the minimum $T$ satisfying:
$$\forall i \in [0, N-1-T]: \mathcal{G}(k_{i+T}) - \mathcal{G}(k_i) \geq U_T \quad (9)$$
where $U_T = \frac{\mathcal{G}(k_{N-1-T}) - \mathcal{G}(k_T)}{L-2}$. Note that $U_T$ monotonically decreases as $T$ increases as the keys are in ascending order.

The naive algorithm to compute the minimum $T$ is simply enumerating the value of $T$ from 0 to $\lceil \frac{N}{3} \rceil$ and checking the validity of

$T$ by Condition (9). However, the time complexity is $O(N^2)$, which is too expensive for index operations.

To solve this problem, we propose the Fastest Minimum Conflict Degree (**FMCD**) algorithm to achieve *linear* complexity for computing the minimum $T$ and the corresponding model $\mathcal{M}$.

In Algorithm 1, we begin by considering the minimum $T$ as 1 (line 3) and setting $U_T$ as $U_1$. Then we traverse the sorted keys in $\mathcal{K}$ and check whether the current $T$ satisfies the condition $\mathcal{G}(k_{i+T}) - \mathcal{G}(k_i) \geq U_T$ (line 5-6). If a key $k_i$ breaks the condition, $T$ is not good. We increment the $T$, update the $U_T$, and continue the check from the failed key $k_i$ (line 9-10). Not until all keys are checked can we jump out of the loop (line 7-8). At this moment, we get the corresponding model's parameters based on the computed conflict degree (line 11-12).

Theorem 3.3 shows the optimum of the **FMCD** algorithm:

THEOREM 3.3. *The $T_0$ returned from Algorithm 1 is precisely the minimum $T$ satisfying Condition* (9).

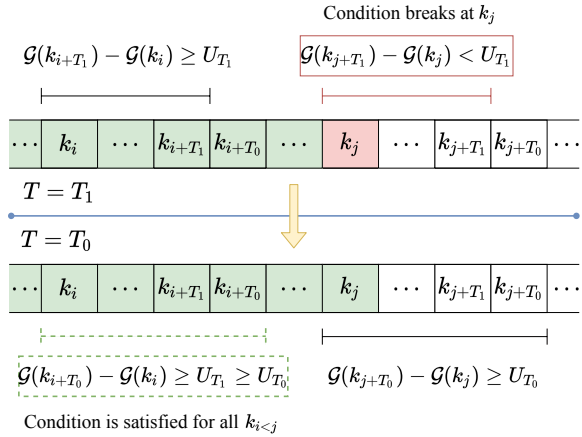

Figure 3: Cases for Proof

SKETCH. Figure 3 shows the relations of $\mathcal{G}(k_{i+T})$, $\mathcal{G}(k_i)$, and $U_T$ when $T$ is changed to $T_0$ for the key $k_j$, which results in the correctness and the minimum of $T_0$. See Appendix [44] A.1 for details. □

**The Complexity:** Finally, we analyze the time and space complexity of Algorithm 1. In fact, the algorithm visits a key only once if the key passes the conditions in line 5. When the check fails, the algorithm increments $T$ in line 9 and visits the key again. Since $T < O(N)$, the number of additional checks due to failed checks is limited to $O(N)$. Therefore, the time complexity of Algorithm 1 is $O(N + \max T) = O(N)$. As for the space complexity, the algorithm needs $O(1)$ space for $U_T$.

## 4 OPERATIONS OF LIPP

In this section, we describe the procedures for index operations and the algorithms to dynamically adjust the index structure.

### 4.1 Lookup and Range Queries

Index read operations include looking up a single key and obtaining a range of keys. We introduce these two types of operations for our proposed LIPP.

**Algorithm 2**: Lookup($\mathcal{T}, k$)

**Input**: $\mathcal{T}$: the LIPP index, $k$: the lookup key
**Output**: *isFound*: indicates whether $k$ is found, $e$: the entry containing $k$ if found

1 **begin**
2   $n \leftarrow$ the root node of $\mathcal{T}$;
3   $e \leftarrow n.\mathcal{E}[n.\mathcal{M}(k)]$;
4   **while** *type*($e$) == NODE **do**
5     $n \leftarrow$ the node pointed to from entry $e$;
6     $e \leftarrow n.\mathcal{E}[n.\mathcal{M}(k)]$;
7   **if** *type*($e$) == DATA **then**
8     $k' \leftarrow$ the key in entry $e$;
9     **if** $k == k'$ **then**
10      **return** $\langle True, e \rangle$;
11  **return** $\langle False, e \rangle$;
12 **end**

*4.1.1 Lookup Queries.* To look up a key, we start at the root node of the index structure, and use the model of the current node to compute the location of the given key in its entry array $\mathcal{E}$. Depending on the entry type, we have different actions.

If the type of the entry is *NODE*, we then follow the pointer to the child node at the next level. Otherwise, we reach the lowest level of the traversal path. If it is a *NULL* entry, the key to lookup does not exist in our index. If it is a *DATA* entry, we should be careful in this case and cannot directly return the current entry. We check whether the search key is the same as the key stored in the entry because different keys can be mapped to the same position. Only when it is consistent can we return the current entry as the lookup result.

The above procedure is listed in Algorithm 2. Note that the returned result contains the entry even when the key is not found. This is used in the insert operation.

It is important to note that there is no need to use extra search steps in each node in our lookup procedure. This is because the model computes the precise position for the key in each node. Therefore, the cost of the lookup query in our index is only $O(h)$, where $h$ is the height of the index tree.

*4.1.2 Range Queries.* A range query is used to find the elements whose keys are in the specified range, which is an important operation in database engines. Given the range $[u, v]$, we first find the position of the start key $u$ by performing the point lookup procedure described above.

Since the index is monotonic, we scan forward until reaching the end key $v$. If we reach the end of the entry array containing the start key before reaching the end key, we traverse back to the previous level and continue the scan. When reaching a *NODE* entry during the scanning process, we follow the pointer to scan the associated child node. We can further use the bit vector to quickly skip over gaps.

One concern is that during the scanning process, we may perform unnecessary comparisons between an entry key and the end key. We address this concern by first locating the positions of the end key $v$ in all levels and then simply visiting the elements up to the computed positions without key comparisons.

**Algorithm 3:** Insert($\mathcal{T}$, $k$, $p$)

**Input**: $\mathcal{T}$: the LIPP index, $k$: the key, $p$: the payload

1 **begin**
2     Entry $e \leftarrow$ **Lookup**($\mathcal{T}$, $k$).$e$;
3     **if** $type(e) ==$ *NULL* **then**
4         $type(e) \leftarrow$ *DATA*;
5         $e \leftarrow \langle k, p \rangle$;
6     **else**
7         $k' \leftarrow$ the key in entry $e$;
8         $n \leftarrow$ a new node trained on $k, k'$;
9         $type(e) \leftarrow$ *NODE*;
10        $e \leftarrow$ the pointer to node $n$;
11     **for** *node $n'$ in the reversed traversal path* **do**
12         **Adjust**($\mathcal{T}$, $n'$);
13 **end**

**Algorithm 4:** Adjust($\mathcal{T}$, $n$)

**Input**: $\mathcal{T}$: the LIPP index, $n$: the node to adjust

1 **begin**
2     $n.element\_num = n.element\_num + 1$;
3     **if** *the insertion conflicts* **then**
4         $n.conflict\_num = n.conflict\_num + 1$;
5     **if** $n.element\_num \geq \beta \cdot n.build\_num$ **and** $\frac{n.conflict\_num}{n.element\_num - n.build\_num} \geq \alpha$ **then**
6         $\mathcal{K} \leftarrow$ the collection of keys contained in the subtree rooted at $n$;
7         $n' \leftarrow$ **BuildPartialTree**($\mathcal{K}$);
8         Replace node $n$ with $n'$ in $\mathcal{T}$;
9 **end**

## 4.2 Index Inserts

Apart from locating the keys, the index should handle insert operations, while maintaining the strict order guarantee for the index. In our case, it is easy to achieve this requirement.

In the insert algorithm, the logic to reach the entry in the final level is the same as in the lookup algorithm described above. When the entry returned by the lookup query is a gap, i.e. *NULL*, we simply insert the new element into this gap. But when a conflict happens, we need to replace the elements with a pointer to the new node containing the new element as well as the original one. At the same time, we change the entry type to *NODE* to indicate the existence of the node.

During the experiments, we find the creation of new nodes with two elements is a common operation. This can exert negative influence on the performance. To improve the performance, we construct our own memory pool for small node allocation and recycling.

After finishing the insert operation, we follow the search traversal path in the opposite direction and judge whether the nodes should trigger the adjustment to keep the index height bounded, as will be detailed in Section 4.3.

## 4.3 Adjustment Strategy on the Model

In this section, we describe the adjustment strategy to keep the tree height bounded. Before doing the actual adjustment, we first update and check the statistics of the nodes in the traversal path after an insert operation. We trigger the adjustment on a chosen node when certain conditions are satisfied. The adjustment procedure is shown in Algorithm 4. To understand the adjustment strategy, we focus on two core issues: When to adjust and How to adjust?

*4.3.1 When to adjust?* In LIPP, we propose two main criteria, as shown in Algorithm 4, to decide whether to trigger adjustment on a node:

Firstly, *the number of inserted elements in the subtree rooted at node n should be at least $\beta$ times as large as the elements in the subtree rooted at n in the last adjustment*. Concretely, the condition can be formulated as the expansion ratio $\frac{n.element\_num}{n.build\_num} \geq \beta$ for node $n$, where $n.build\_num$ is the number of elements specified to $n$ by the adjustment (line 13 in Algorithm 5) and $n.element\_num$ is incremented for each insertion. $\beta$ is set to 2 by default. The default value is derived from the logarithm methods of PGM, i.e., always triggering the merge process for a series of indexes when the inserted elements is twice as much as elements contained in the largest index. This criterion effectively reduces the frequency of adjustments and leads to low adjustment complexity even in the worst cases (cf. Section 5).

Secondly, *the conflict ratio between the number of conflicts and insertions in the subtree rooted at node n exceeds a given threshold $\alpha$*, i.e. $\frac{n.conflict\_num}{n.element\_num - n.build\_num} \geq \alpha$. This is because conflicts result in the creation of new small nodes and possibly increase the tree height. Consequently, conflicts tend to hurt the performance for both lookup and insert operations. Thus, we should trigger adjustment if we have seen too many conflicts in the subtree rooted at $n$. By default, we set the threshold $\alpha = 0.1$ to achieve a better balanced tree. To further choose proper parameters, we conduct the parameter analysis for $\alpha$ and $\beta$ in Appendix [44] B.2.

In addition, we observe that **adjustment on a node with either too many or too few elements downgrades the performance.**

For the first observation, the extreme case is that when the root node is triggered to adjust, all elements should participate in the adjustment, which may not be acceptable for a very large tree. Therefore, we should restrict the number of elements participating in the adjustment. When the node size is close to the predefined threshold (e.g. 16MB) which means that the node has already contained enough entries and elements, the current node should be fixed and not participate the adjustment afterwards. As for the second observation, when the number of elements of a node is small, a few insertions will trigger adjustment in the early stage. To eliminate such case, the nodes with less than 64 elements are required not to trigger the adjustment.

*4.3.2 How to adjust?* In Algorithm 4, if a node $n$ satisfies the above conditions, the adjustment is triggered. First of all, we collect all elements in the subtree rooted at node $n$ by sequential traversal (line 6). The keys of the collected elements are in order because the tree is sorted. Then, we build a partial tree structure on the elements (line 7). Finally, we update the pointer of the original node to point to the root of the new tree structure (line 8). Figure 4 illustrates the node adjustment procedure. The main procedure of adjustment is to build a proper tree structure based on the collection of keys. Utilizing the properties of models and conflicts in Section 3.3, the optimal tree structure should consist of nodes with best models having the minimum conflict degree. The resulting models distribute the elements in the entry array to minimize conflicts, thereby

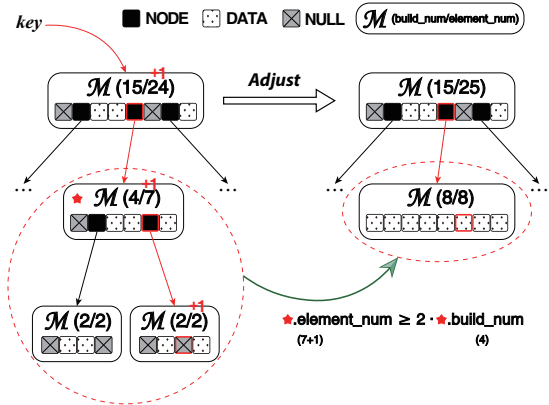

Figure 4: Node Adjustment

reducing the number of elements assigned to the next level and limiting the tree height.

In Algorithm 5, we use the **FMCD** algorithm as described in Section 3.4 to compute the best model for each node. Given a collection $\mathcal{K}$ of keys, we create a new node $n$ with $L$ entries, where $L$ is $\delta$ (e.g., $\delta = 2$) times of the number of keys to preserve enough gaps for future insertions after the adjustment (line 2). $L$ is also bounded by the pre-defined node size upper bound, i.e., $L \leq 16MB/16B = 1M$. In fact, $16MB$ is set from the max node size in ALEX, and $L = 1M$ in our experience always produces good results and does not need to be tuned. Next, we obtain the model for the new node $n$ on the given keys by **FMCD** (line 3). After that, we insert every key into the entry given by the model. When there is a conflict, i.e., more than one element are mapped to a given position $l$, we collect conflicting elements into $\mathcal{K}_l$, recursively build the partial tree on $\mathcal{K}_l$ to create a subtree, and set the entry at position $l$ to point to the child tree (line 9-12). For a non-conflicting element, we simply insert it to the entry predicted by the model (line 6-8). Finally, we initialize the basic statistics for the new node, which will be used by the next adjustment (line 13-14). The returned node $n$ is the node to replace the original node in LIPP.

We will analyze the impact of adjustment on lookup and insert operations in Section 5.

### 4.4 More Operations

**Bulkload:** The bulkload operation follows the same procedure as the partial tree building in algorithm 5. The returned result is the root node.

**Delete:** Delete is supported by looking up the entry of the element and marking the type of that entry as *NULL*. If a future lookup query for this deleted key is executed, it just traverses to a *NULL* entry, which indicates non-existence of the key. It may be argued that nodes should be either contracted or merged after entry deletions. Since real-world datasets tend to increase over time, we believe retaining the gaps generated by deletions is beneficial to subsequent insertions. Thus, for simplicity, we do not modify tree structures for deletions.

**Update:** There are two types of updates. One is to modify the key, and the other is only to modify the payload. The former is implemented by combining a delete operation with an insert operation, while the latter is supported by looking up the key and over-writing the payload.

---

**Algorithm 5**: BuildPartialTree($\mathcal{K}$)

**Input**: $\mathcal{K}$: the collection of keys to adjust
**Output**: $n$: the new root node of partial tree

1 **begin**
2    $n \leftarrow$ new node with $L = \max(1M, \delta \cdot |\mathcal{K}|)$ entries;
3    $n.\mathcal{M} \leftarrow$ **FMCD**($\mathcal{K}, L$).$\mathcal{M}$;
4    **for** $l \in [0, L]$ **do**
5      $\mathcal{K}_l = \{k \in \mathcal{K} | n.\mathcal{M}(k) == l\}$;
6      **if** $|\mathcal{K}_l| == 1$ **then**
7        $type(n.\mathcal{E}[l]) \leftarrow DATA$;
8        $n.\mathcal{E}[l] \leftarrow k \in \mathcal{K}_l$;
9      **else if** $|\mathcal{K}_l| > 1$ **then**
10        $type(n.\mathcal{E}[l]) \leftarrow NODE$;
11        $n' \leftarrow$ **BuildPartialTree**($\mathcal{K}_l$);
12        $n.\mathcal{E}[l] \leftarrow$ the pointer to node $n'$;
13    $n.element\_num \leftarrow |\mathcal{K}|$; $n.build\_num \leftarrow |\mathcal{K}|$;
14    $n.conflict\_num \leftarrow 0$;
15    **return** $n$;
16 **end**

---

**Overflow Insert and Lookup:** An overflow operation means that a key exceeds the existing key space in the tree. For instance, the append-only insert workload may cause consecutive overflow inserts. In LIPP, the overflow insert and lookup can be optimized by maintaining two temporary sorted buffers for the right-most and the left-most entries and triggering adjustment when either buffer is full. Note that this design will not affect the normal insert and lookup operations because normal operations do not visit the two extra buffers. We use binary search in the buffers to improve search performance for overflow operations.

## 5 ANALYSIS

In this section, we provide complexity analysis for both lookup and insert operations, which is already shown in Table 1. Other operations can be analyzed in the similar way.

### 5.1 Tree Height and Lookup Analysis

In LIPP, the lookup traverses along a path from root to the final entry without extra search steps. Therefore, the lookup cost is only related to the height of the index structure.

A tree index can be either built from bulkload operations, i.e., an adjustment on the root node, or result from a number of random insertions. We first analyze the tree height for the former case: Let $m$ be the minimum fanout of an index, i.e., $\min \lceil \frac{N}{T} \rceil$ among all nodes. In fact, $m \geq 3$ since $T \leq \lceil \frac{N}{3} \rceil$, which indicates that at least $\lceil \frac{N}{T} \rceil \geq 3$ positions each contains up to $T$ entries for any given $N$ elements.

**Theorem 5.1.** *The height of a LIPP index with $N$ elements built from adjustment is at most $O(\log N)$.*

**Proof.** Even in the extreme case, for each level of the index structure, there are at least $m$ branches, each of which contains at most $\frac{N}{m}$ elements. Thus, the height of index is at most $O(\log_m N)$. □

Now we pay more attention to the height of the index built from scratch in Theorem 5.2, which is a more common way for index construction in practice.

Theorem 5.2. *The height of a LIPP index with N elements is at most $O(\log N)$.*

Sketch. The core idea is to find the recursive relation between the number of elements existed in the parent and the child node. Refer to Appendix [44] A.2 for details. □

Finally, as the complexity of lookup operations depends only on the tree height, it is obvious that the complexity of a lookup operation is $O(\log N)$, even in the worst cases.

In the average case, it is obvious that LIPP has the complexity $O(\log_m N)$ for lookup operations when the fanout of nodes is at least $m$. Compared with PGM, which takes $O(\log_m^2 N)$, and ALEX, which takes $O(\log_m N + \log_2 m)$, LIPP achieves the best complexity for lookup operations.

### 5.2 Insert Analysis

Though the cost of a normal insertion also depends on the tree height, we need to concentrate on the cost of adjustment, which incurs extra overhead for insertions. Since adjustment is not triggered by each insertion, it is reasonable to employ the amortized analysis for insert operations.

Firstly, we analyze the adjustment cost in the worst case.

Theorem 5.3. *The adjustment on a node with N elements costs at most $O(N \cdot \log N)$.*

Proof. With $N$ elements, obtaining the best model with the minimum conflict degree costs $O(N)$. Meanwhile, the height for the adjusted tree is $O(\log N)$. In the worst case, each level would cost $O(N)$ to train models. Then the total cost would be at most $O(N \cdot \log N)$. □

Then, for the LIPP index with size $N$ built from the insertions, we only pay amortized $O(\log^2 N)$ for each insertion:

Theorem 5.4. *The amortized cost for insert operations is at most $O(\log^2 N)$.*

Sketch. We use the accounting method [10] for the analysis. The core idea is to save extra $O(\log N)$ credits for each node along the traversal path of insertions and to consume credits for the adjustment. Please refer to Appendix [44] A.3 for details. □

Insertions with amortized complexity $O(\log^2 N)$ for LIPP seems worse than B+Tree. But compared with other learned indexes, we still achieve better complexity. In fact, ALEX costs $O(m)$ for shifting the elements in the extreme cases where $m$ is the average fan-out of a node. Besides, our performance is analyzed in the worse case rather than average case. In most cases, an insertion only needs to traverse to the *NULL* entry and add the element, which costs only $O(\log_m N)$. Moreover, experimental results in Section 6 show that insertions behave well in practice.

## 6 EVALUATION
### 6.1 Experiment Setup

**Datasets:** We evaluate our method using four popular benchmarks mainly used in [11], with the detailed statistics shown in Table 2: (1) Longitudes (LTD) consists of the longitudes of locations around the world from Open Street Maps[1]. (2) Longlat (LLT) consists of compound keys which combine the longitudes and latitudes from Open Street Maps by applying transformation to these pairs, whose distribution is severely non-linear. (3) Lognormal (LGN) is generated artificially according to the log-normal distribution[2]. Besides, the values are multiplied by $10^9$ and rounded down to the nearest integers. (4) YCSB is also artificially generated, representing the user IDs from YCSB benchmark[3]. These values are uniformaly distributed across int64 domain. No duplicated elements are contained in these datasets unless otherwise stated. We also randomly shuffle these four datasets to simulate the real-world scenarios, which follows the settings mentioned in paper [11].

Table 2: Statistics of Datasets

|              | LTD    | LLT    | LGN   | YCSB  |
|--------------|--------|--------|-------|-------|
| Num Keys     | 1B     | 200M   | 190M  | 200M  |
| Key Type     | double | double | int64 | int64 |
| Payload Size | 8B     | 8B     | 8B    | 80B   |
| Total Size   | 16G    | 3.2G   | 3.04G | 17.6G |

**Baselines:** We compare LIPP with existing state-of-the-art baselines: (1) Standard B+Tree, as implemented in the STX B+Tree [5]. It is the fastest implementation for all in-memory B+Tree structures over many common operations. (2) Adaptive Radix Tree(ART) [26], a trie optimized for main memoring indexing and adapted to the data. (3) ALEX [11], an in-memory, updateble learned index, which utilizes the gapped array to accommodate elements with exponential search step. (4) PGM [12], a fully-dynamic compressed learned index with provable worst-case bounds. (5) Learned Index [22], using a two-level recursive model index with linear model at each node and binary search steps which only support lookup operations. (6) BwTree [27], a general purpose, concurrent and lock-free B+-Tree index. These source codes are publicly available. We download and run their codes following the guidelines specified in the papers.

**Workloads:** The primary metric is the average throughput for LIPP compared with other methods. To demonstrate the performance over different index operations, we evaluate the throughput on different types of workloads: (1) The Read-Only workload, which performs lookup operations on the indexes built from bulkloading 100M randomly selected keys. (2) The Read-Heavy workload, which contains 33% inserts to put elements into indexes and 67% reads to randomly lookup elements. (3) The Write-Heavy workload, which contains 67% inserts and 33% lookups. (4) The Write-Only workload, which consists of only insert operations. The workloads (2) through (4) are tested on an empty index. Besides, the keys to lookup are selected randomly from the set of existing keys in the index, and the keys to insert are randomly chosen from non-existed ones. We run the workloads on different indexes within 100M operations for five times and obtain their throughput of operations (either lookup or insert) finally.

**Environment:** We implement LIPP in C++ and compile it with GCC 9.0.1 in O3 optimization mode. All experiments are conducted on an Ubuntu 18.04 Linux machine with 4.0 GHz Intel Core i7 (4 cores and 8 threads) and 32GB memory, only using a single thread.

---

[1]https://registry.opendata.aws/osm
[2]https://en.wikipedia.org/wiki/Log-normal_distribution
[3]https://github.com/brianfrankcooper/YCSB.git

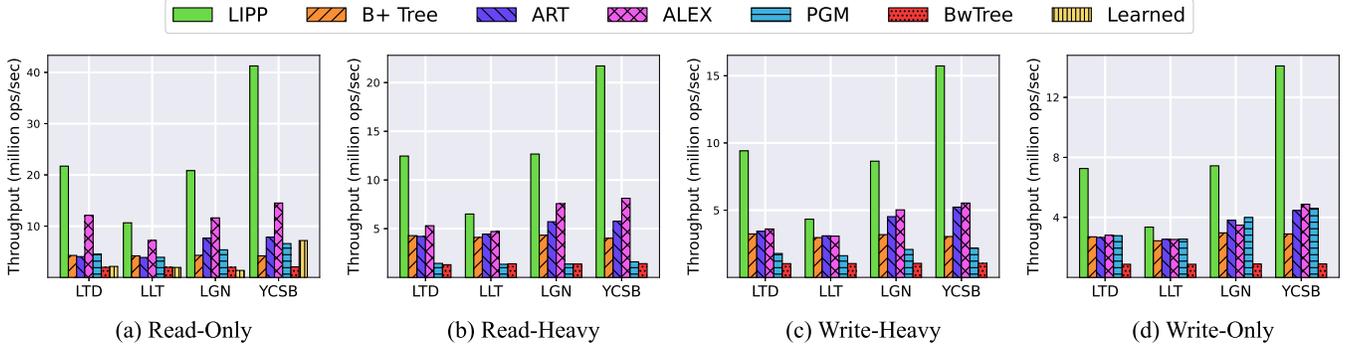

Figure 5: Throughput: Comparisons with State-of-the-art Methods

## 6.2 Compare with State-of-the-art Methods

Next we compare LIPP with state-of-the-art methods. The results are shown in Figure 5. To sum up, our approach achieves promising results and consistently beats those state-of-the-art methods on different types of workloads.

### 6.2.1 Read-Only Workloads.
For read-only workloads, LIPP achieves up to 2.8x, 6.3x, 15.3x, 9.8x throughput than ALEX, PGM, Learned Index and B+Tree, respectively. Compared with Learned Index, ALEX and PGM, LIPP is able to eliminate the in-node search step while keeping the tree height bounded. Therefore, it incurs less computation overhead to locate the correct position. Besides, for those traditional indexes B+Tree and ART, LIPP also has larger nodes to accommodate many more elements on one level, and thus has less tree height. Thus, the traversal path of LIPP is significantly shorter than those of traditional indexes, resulting in better performance. BwTree behaves even much worse than B+Tree since it costs more time on the infrastructure to allow concurrent operations.

Meanwhile, on the LLT dataset, the throughput for LIPP is only 1.47x than ALEX. But we see that ALEX and PGM even behave worse or similar than Learned Index on this dataset. The reason is that LLT is highly non-uniform, which makes it difficult for learned model to depict overall distributions and thus the tree height of indexes increases. As a result, the traversal on tree structure dominates the overall performance and the performance of all learned index structures tends to be similar.

### 6.2.2 Read-Write workloads.
For read-write workloads, LIPP again achieves better performance than existed approaches with comparable index size. Since Learned Index does not support any update operations, we exclude it from the comparisons here. Figure 5 indicates that LIPP beats ALEX, PGM, B+Tree, ART with 2.9x, 13.5x, 5.4x, 3.7x higher throughput on performance.

We observe that the performance of all methods decreases with the increasing portion of insert operations. The reason is that adjustments are required to deal with changes in the index structure caused by insert operations. Compared with the long distance shifting of ALEX, the new node creation of LIPP is more light-weighted. Since the adjustment helps to bound the tree height, LIPP still has short traversal path for insert operations. Therefore, LIPP better handles the write operations than ALEX. Besides, PGM performs even worse in read-heavy workloads since PGM needs $O(\log N)$ trees to support updating and each lookup operation should perform searching in all these trees. Since the traversal cost is still the main bottleneck for traditional indexes, LIPP obviously beats B+Tree and ART.

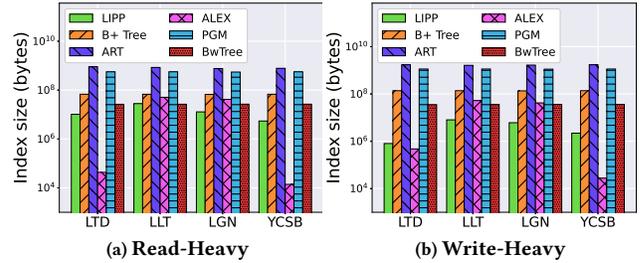

Figure 6: Index Size

### 6.2.3 Index Size.
Due to space limitation, we just display the index size on read-heavy and write-heavy workloads in Figure 6. In general, the index size of LIPP is not sensitive to the distribution of datasets, and is comparable or less than that of ALEX and Learned Index in read-write workload except for YCSB. The reason is that in YCSB, the model can fit the distribution almost perfectly and therefore ALEX can better trade the search performance for reducing the index size by holding the inaccurate elements in wrong positions. At the same time, LIPP needs extra small nodes in the lowest level to hold conflicting elements in local areas. Besides, we also observe that PGM has larger index size because it utilizes the logarithmic method to support insert and therefore requires to store a series of index trees.

Nevertheless, we point out that the space overhead for the raw data will overshadow that of index. For example, the size of YCSB benchmark is 17.6 GB while the index size is not larger than 500 MB even for the worst method. Therefore, the index size will not be the bottleneck of memory usage and the index overhead of LIPP is acceptable.

### 6.2.4 Range Query.
The range query operation can be implemented by a simple lookup operation on the start key and a scan for the following elements. We restrict the number of keys for scanning to less than 100 for range queries, as mentioned in paper [11]. For read-only workload, the performance of range query is comparable with ALEX and B+Tree as shown in Figure 7, just keeping its advantages up to 1.5x performance gain. The reason is that as scan time begins to dominate the overall query time, the speedup of LIPP on lookups becomes less apparent. Besides, LIPP needs to distinguish the different entry types and apply different actions during the scan, which also slightly influences the performance of range query. It is also worth mentioning that ART fail to support range due to its implementation. Since PGM needs to insert elements in a series of indexes, it requires to scan all components to support the range query in PGM. Thus the query time of PGM is orders of magnitudes slower than LIPP and thus is ignored here.

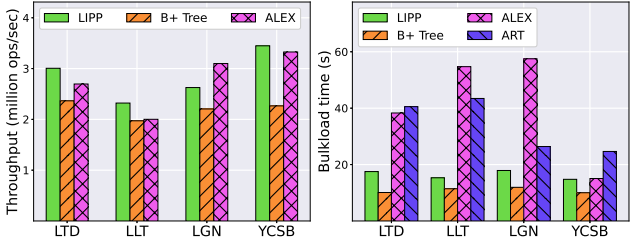

Figure 7: Range Query    Figure 8: BulkLoad

*6.2.5 Bulkload.* In Figure 8, we demonstrate the time to load 100M elements at once to build the index. Among all learned indexes, LIPP only requires less than half of the time to bulkload compared with ALEX and PGM. Actually, LIPP only needs to create new nodes for conflicting elements instead of shifting the elements to preserve a gap for ALEX. Also, PGM builds its index in a bottom-up manner and results in more segments with specified error bound.

In addition, LIPP is only slightly slower than B+Tree on bulkload since LIPP needs to create small nodes for conflicting elements. But LIPP can make up for its little slower bulkload time. Moreover, after we further insert 100K records after bulkload, the performance of insert operation in B+Tree decreases sharply due to the aging problem.

### 6.3 Detailed Performance Study

| R/W  | LIPP    | ALEX     | PGM    | Learned | B+Tree    |
|------|---------|----------|--------|---------|-----------|
| LTD  | 3.6/4.5 | 7.5/13.0 | 19.0/- | 20.7/-  | 56.5/57.6 |
| LLT  | 4.4/6.8 | 9.2/17.4 | 19.0/- | 36.0/-  | 56.8/57.8 |
| LGN  | 3.5/4.1 | 8.1/15.5 | 18.0/- | 28.8/-  | 57.9/57.7 |
| YCSB | 3.1/3.1 | 9.1/10.7 | 17.0/- | 12.6/-  | 57.0/57.8 |

Table 3: The Average Number of Memory Accesses

Next we investigate how the contribution that each proposed technique makes to the overall performance. In Table 3, we display the average number of memory accesses for lookup and insert operations on read-only and write-only workloads, separately represented by two values in each cell. For insert operations, we only count the most common cases without structure modifications, which reveals the length of common paths accessed by insertions. Besides, since PGM uses logarithm methods to support insertion, its memory access is meaningless for single insert operation. ART just behaves similar to B+Tree. Thus we ignore them here.

The results in Table 3 show that LIPP has the significantly fewer memory accesses compared with other baselines. In fact, the number of memory accesses is at least 3 for the indexes with two or more levels. That is, one access to get the entry in root model, one access to extract the pointer and obtain the child model, and one access to locate the entry of the lookup key.

For lookup operations, LIPP eliminates the "last-mile" search of ALEX and Learned Index, removes redundant lookups in different components of PGM and also reduces the traversal path over index trees compared with B+Tree and ART. Compared with LIPP, ALEX requires extra memory accesses to finish the exponential search to correct the mispredications in leaf nodes while PGM needs to find keys in a series of subtrees and thus incurs more memory accesses. Moreover, since B+Tree index has many more levels, the lookup operations need to traverse the overall levels and figure out the internal node by binary searching, which thus costs more overhead as shown in the table. All improvements are reflected by the small number of memory accesses. For insert operations, the adjustment operation of LIPP is much cheaper than shift in ALEX and split in B+Tree. In fact, ALEX aims to use a gapped array to reduce the number of shifting elements to $O(\log m)$ with high probabilities, but still incurs huge costs for each insertion and introduces the overhead to maintain the structure of the gapped array.

To further display the internal properties of our method, we conduct the experiments of variance analysis on the write-heavy workload. In one run, we collect the latency (execution time) of each operation, sort the latencies in the ascending order, and report 99-th percentile latency in Figure 9a. LIPP has relative small latency compared with ALEX and B+Tree. This is due to that LIPP has fewer structure modifications, which is shown in Table 3a. Therefore, most of our operations are handled in the common cases without incurring too much overhead. Besides, LIPP utilizes the fewer memory accesses to finish these operations, which causes the lower latency. As for ALEX, its structure modifications are more common than LIPP [11]. While PGM uses logarithm methods to support insertions, it will frequently incur merge process during insert and lookup operations, which causes the larger 99-th percentile latency.

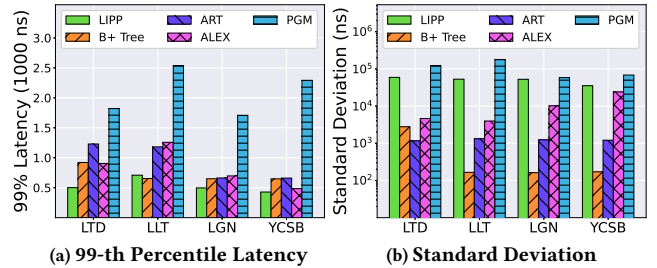

(a) 99-th Percentile Latency    (b) Standard Deviation

Figure 9: Variance of Operations

We also report the standard deviation of the operation time in Figure 9b. Though LIPP seems not outstanding, the standard deviation of our method is still comparable with those learned indexes, i.e., ALEX and PGM. Actually, the structure adjustment of the traditional indexes only involves a small amount of data. Thus, the latency of each operation for B+Tree and ART is relatively uniform. However, the structure modifications of learned indexes involve the larger number of keys and values and cost much more time to finish, which causes the inevitable higher standard deviation. In a word, the high standard deviation of our method is acceptable and reasonable.

Besides, we report the number of adjustments, as well as the ratio of time spent on adjustment to the overall running time. The results in Table 3a illustrate the rarity of structure modifications on LTD, LGN and YCSB and the results in Table 3b show that the time used in adjustments is relatively small compared with the running time, which thus is not the main bottleneck of LIPP. Besides, we find that a single adjustment may spend much more time than an insert operation without adjustment triggered. However, it won't be a severe case since the adjustment is rarity and more than 99% of insert operations do not trigger adjustments. And in Appendix [44] B.3, the ratio of time spent on structure modifications for ALEX and adjustments for LIPP are comparable, which also alleviates the severity of the above problem.

| (a) # Adjustment | | | | (b) % Adjustment | | | |
|---|---|---|---|---|---|---|---|
| | RH | WH | WO | | RH | WH | WO |
| LTD | 27 | 32 | 32 | LTD | 25% | 33% | 31% |
| LLT | 2448 | 10887 | 20594 | LLT | 13% | 17% | 17% |
| LGN | 22 | 24 | 24 | LGN | 23% | 26% | 22% |
| YCSB | 11 | 11 | 11 | YCSB | 23% | 21% | 17% |

Table 4: The Number and Time Ratio of Adjustments

Moreover, we explain the behaviour of our method on LLT dataset. Since LLT is highly non-uniform and its distribution is roughly jagged, LIPP utilizes many more adjustments to adapt this irregular dataset. But the most of the adjustments only involve a very small part of the data, which thus do not cost too much time. As we can see in Table 3b, the time for adjustment caused on LLT is still small compared to the overall running time, whose ratio is only 13%, 17%, 17% on read-heavy, write-heavy and write-only workloads.

## 6.4 Scalability with Data Size

Here we evaluate the scalability of LIPP with the increasing volume of data. We run both the read-heavy and write-heavy workloads on the LTD dataset with the varying number of keys in the beginning, instead of building the index from scratch.

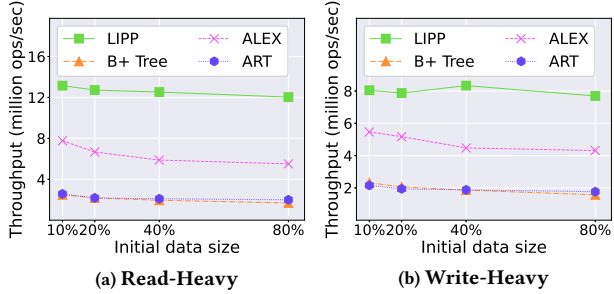

(a) Read-Heavy  (b) Write-Heavy

Figure 10: Data Scalability

Figure 10 shows the results for both read-heavy and write-heavy workloads. We see that LIPP beats ALEX and B+Tree with the increasing data size. As the number of elements increases, the throughput of LIPP slows down, but in a low rate since the tree height is restricted by the FMCD strategy in adjustment. We also point out that when the data volume is 40% in Figure 10b, the abnormal increasing of LIPP is caused by the different tree structure of indexes after bulkload on different number of keys.

## 6.5 The Effect of Adjustment Strategy

To show the effectiveness of our proposed algorithm for adjustment, we compare LIPP using the default adjustment strategy (FMCD) with LIPP simply using Linear Regression (LR) to obtain the model parameters $\mathcal{M}.A$ and $\mathcal{M}.b$. Figure 11a shows the results, where FMCD consistently achieves better than LR, even on YCSB that is totally uniform. FMCD distributes different keys to make as few conflicting elements as possible, while LR only tries to approximate the key and position with straight lines. Moreover, the low complexity of FMCD restricts the cost of reconstruction not higher than the performance improved by this strategy, making FMCD more competitive. To further demonstrate the dominance of our adjustment strategy, we apply LR and FMCD on $1M$ random keys from different datasets to get two models, respectively. We then obtain the conflict degree of these two models in Table 11b. The results show that FMCD has many fewer conflict degrees than LR consistently, which would have the fewer layers for all keys in average and thus improve the overall performance. Actually, LR is sensitive to the elements, where a drift on the last point results in larger slope and thus causes more conflicts. But FMCD strategy is more robust to the distribution, and tries its best to limit the conflict degree to a very small number.

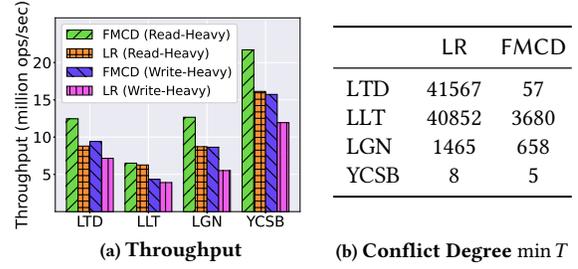

(a) Throughput  (b) Conflict Degree min $T$

| | LR | FMCD |
|---|---|---|
| LTD | 41567 | 57 |
| LLT | 40852 | 3680 |
| LGN | 1465 | 658 |
| YCSB | 8 | 5 |

Figure 11: Effect of Adjustment Strategy

## 7 DISCUSSION

### 7.1 Concurrent Operations

Among past years, researchers have extensively studied how to implement a concurrency control protocol in B+Tree or other trees [3, 24, 36]. In fact, LIPP can be considered as a tree and the adjustment on LIPP can be viewed as a type of structure modification, similar to the split or merge process in B+Tree. Thus, traditional ways to support concurrency can be used in our method. We could apply the latch coupling [3] protocol on LIPP, which needs read/write locks. For lookup operations, we repeatedly acquire the read lock on child and then unlock parent along the traversal path. For insert operations, we traverse index and repeatedly obtain the write lock on a child. Once the child is locked, we check if it is safe, i.e., this node will not trigger the adjustment after inserted. If it is safe, we release locks on ancestors.

Besides, LIPP's unique advantage of precise positions reduces concurrency contentions compared to B+Tree and other learned indexes. In B+Tree or ALEX, insertions without structure modifications may cause elements to shift in the leaf nodes. Consequently, lookup and insert operations in the same leaf node may still incur contentions even though these two operations on different positions. However, LIPP effectively eliminates such unnecessary contentions since insert operations with no adjustment do not affect lookup operations in other positions. Thus, we can remove mechanisms for those contentions in concurrency control protocols.

## 7.2 New Hardware Accelerators

New hardware accelerators, such as GPU/TPUs, will make our method even more valuable. At the same time, those new hardwares have their own challenges, most importantly the high invocation latency. It still requires 2-3 micro-seconds to invoke any operation on them. However, with the expansion of video memory and the development of direct NVM access from GPU [35], it is reasonable to assume that probably all learned indexes will fit on the GPU. Besides, the integration of machine learning accelerators with the CPU is getting better and with techniques like batching requests, the cost of invocation can be amortized, so that we do not believe the invocation latency is a real obstacle.

Another challenge is under-utilization brought by the totally different computing models of new hardwares. We need to fully utilize the thousands of available GPU cores, which requires eliminating or at least minimizing required communication between GPU threads and branch divergence within a SIMD instruction. Therefore, we may need to change the underlying data layouts to avoid the contention as much as possible. Moreover, the synchronization and communication in GPU are far more heavy-weighted and complicated than that in CPU, which may require new concurrency control protocols.

## 8 RELATED WORK

**Learned Index** With development of machine learning, a new family of index structures, such as A-Trees [13] and Learned Index [22], are introduced to learn the underlying data distribution to index data items. This work inspires a series of work. FITing [14] replaces the leaf nodes of a B+Tree with linear models to compress index and maintain operations' performance, which has restricted guarantees for prediction error. PGM [12] extends the idea from FITing, replaces the non-leaf node with linear models also and provides an optimal way to obtain piece-wise linear model under their requirements. Both of PGM and FITing do support insertions but just in naive and direct ways such as extra buffer and merge, which thus are not their main contributions. ALEX [11], indeed, supports the insert operations by utilizing almost fixed non-leaf nodes as route to and applying exponential search on the gapped array in their leaf nodes. However, these indexes just learn the approximate position for elements and need extra search step to correct the predictions. Therefore, they introduce another set of space-time trade-offs between prediction error and tree size. LIPP is totally different, which introduces new structure to completely eliminate the inaccuracy caused by learned model predictions. Besides, Nathan et al. [37] put attention on the multi-dimensional in-memory learned index. Kipf et al. [21] focus on the constructions of learned index in single pass. Tang et al. [42] devise a scalable learned index for multicore data storage. All of them are different from our scope.

**Database Index** Meanwhile, various indexes are proposed over the last decades. Traditionally, B+Tree [2] and its variants are originally proposed for disk settings. However, realizing the importance of cache utilization for memory indexes, several cache conscious B+Trees [15] have entered the fields of researchers' vision. Among of them, Adaptive Radix Tree [26] integrates the B+Tree with Tries to reduce cache misses, while CSS-trees [40] restrict index node size to fit into CPU cache line and use arithmetic operations to eliminate the usage of pointers in index nodes. Instead, pB+-tree [8] uses larger index nodes but mainly relies on instructions to prefetch the necessary information in advance. FAST [18] exploits the usage of SIMD instructions to further optimize searches within index nodes for cache performance. Meanwhile, Blink-Tree [28] provides a concurrent tree restructuring mechanism for handling underflow nodes as well as overflow nodes. Except the series of B+Tree, the choices of index for databases also have many candidates for specific purposes, such as T-trees [25], red-black trees [7], LSM tree [38], and more recently, HOT [6]. Besides, some researchers put their attentions on new hardware platforms, such as Multi-Core Chips [27], Non-Volatile Memory [9, 29], Hardware Transactional Memory [41], Solid-State Disk [17], Graphic Processing Units [4, 19], etc.

**Machine Learning for Database Systems** Machine Learning has been proved its power in different database related fields. Pavlo et al. [39] construct "self-driving" database systems which can automatically optimize themselves without human intervention to cope with complicated query workload and data characteristics. Meanwhile, Ma et al. [30] follow the ideas and further design a query workload forecasting framework in autonomous DBMS. Moreover, Aken et al. [1] propose an automatically database tuning framework boosted by machine learning techniques. More researchers also put focus on combining learning techniques with traditional SQL optimization in different aspects. Marcus et al. [32–34] tackle the problem of query optimization in database system. Meanwhile, Wu et al. [43] also learn cardinality models for overlapping subgraph templates on the shared cloud workloads. Yang et al. [47] study the problem of selectivity estimation using deep learning techniques, which approximate the joint data distribution without any independence assumptions. Also, Kristo et al. [23] introduce a new type of distribution sort that leverages a learned model of the empirical CDF of the data. In addition, Zhang et al. [48] propose a learned scheduler that leverages overlapping data reads among incoming queries and learns a scheduling strategy that improves cache hits. Besides, Yang et al. [46] adopt machine learning techniques for data partitioning to improve storage issues of database system. And Yang et al. [45] utilize a machine learning method to predict hot records in an LSM-tree storage engine and prefetch them into the cache. Hsu et al. [16] devise new algorithms which learn relevant patterns for data streams and use them to improve its frequency estimates.

## 9 CONCLUSION AND FUTURE WORK

In this paper, we introduce LIPP, a brand new learned index structure to efficiently support a full set of index operations. We address the bottleneck of previous learned index structures by precisely predicting the position of a search key and thus eliminating the last-mile searches within leaf nodes. To this end, we provide a novel metric to decide the layout of tree nodes and propose a dynamic adjustment strategy to tightly bound the height of the tree. Experimental results on popular datasets demonstrate the superiority of our proposed method on a variety of workloads.

For the future work, we plan to explore the ways to automatically and effectively select the monotonic functions to make LIPP more robust to datasets with different distributions. Besides, it is also interesting to investigate how to implement LIPP into real world relational DBMS with concurrency control and how to adjust LIPP for new hardwares.

# A PROOFS

## A.1 Proof of Theorem 3.3

**The Correctness of $T_0$:** We ensure that $T_0$ obtained in Algorithm 1 satisfies Equation (9). In Figure 3, we show the time point when $T$ is changed to $T_0$ on the key $k_j$. We then prove $\forall i \in [0, N-1-T_0]$, $\mathcal{G}(k_{i+T_0}) - \mathcal{G}(k_i) \geq U_{T_0}$. Obviously, the constraint is established for any $i \geq j$ since $T_0$ has passed this check in line 5 of Algorithm 4. For $i < j$, as shown in Figure 3, the algorithm has checked a $T_1 < T_0$ and $T_1$ satisfies $\mathcal{G}(k_{i+T_1}) - \mathcal{G}(k_i) \geq U_{T_1}$. Because of the monotonicity of the kernel function and $U_T$, we have:

$$\mathcal{G}(k_{i+T_0}) - \mathcal{G}(k_i) \geq \mathcal{G}(k_{i+T_1}) - \mathcal{G}(k_i) \geq U_{T_1} \geq U_{T_0} \quad (10)$$

Thus, $T_0$ is a correct solution for Condition (9). □

**The Minimum of $T_0$:** Suppose there exists a smaller $T_1 < T_0$ satisfying Condition (9). However, since $T_1$ is not the result returned from Algorithm 1, there must exist a time point when $T = T_1$ fails to pass the check in line 5 and causes the increment of $T$ in line 9. In Figure 3, we can find keys $k_j$ and $k_{j+T_1}$ where $\mathcal{G}(k_{j+T_1}) - \mathcal{G}(k_j) < U_{T_1}$, which means $T_1$ cannot be the solution of Condition (9). This is a contradiction. □

## A.2 Proof of Theorem 5.2

Assume the latest adjustment is triggered when $\mu < N$ elements have been inserted. Then we have $\frac{N}{\beta} < \mu < N$. From theorem 5.1, the maximum number of entries in a subtree of the index after this adjustment is at most $\frac{\mu}{m}$. Then there remain $N - \mu$ elements to be inserted. Even if those remaining elements are inserted to one child, the maximum elements contained in one node are limited to $N - \mu + \frac{\mu}{m} = N - \frac{(m-1)\cdot\mu}{m} < (1 - \frac{m-1}{m\cdot\beta}) \cdot N$. In other words, when a node contains $N$ elements, the maximum number of elements in its subtree is at most $(\frac{m\cdot\beta - (m-1)}{m\cdot\beta}) \cdot N$. Therefore, the height of the index with $N$ elements is at most $O(\log_{\frac{m\cdot\beta}{m\cdot\beta-(m-1)}} N)$, which is $O(\log N)$ since $\frac{m\cdot\beta}{m\cdot\beta-(m-1)} > 1$ when $m \geq 3$ and $\beta > 1$. □

## A.3 Proof of Theorem 5.4

We use the accounting method [10] for this analysis. The core idea is to save extra $O(\log N)$ credits for each node along the traversal path accessed by insertions and consume credits for the adjustment.

Without adjustment, the cost for each insertion is obviously $O(\log N)$ since we simply traverse to *NULL* or *DATA* entry and put elements on *NULL* entry or create new nodes on *DATA* entry. In fact, the traversal path of insertions contains at most $O(\log N)$ nodes. For each node in this traversal path, we store $\frac{\beta}{\beta-1} \cdot O(\log N)$ credits as shown in Figure 12. Therefore, we need $\frac{\beta}{\beta-1} \cdot O(\log N) \cdot O(\log N) = O(\log^2 N)$ credits in total for each insertion.

Next, we display how to use these extra credits to pay the cost of the adjustment. Once a node with $\mu$ elements is adjusted, then the credit on this node is cleared to 0. Only after $(\beta - 1) \cdot \mu$ elements are inserted to this node can we adjust this node. And now we have already collected $(\beta - 1) \cdot \mu \cdot \frac{\beta}{\beta-1} \cdot O(\log N) = O(\beta \cdot \mu \cdot \log N) \geq O(\beta \cdot \mu \cdot \log(\beta \cdot \mu))$, which is enough to pay the current adjustment. Therefore, as shown in Figure 12, the adjustment triggered from a node simply consumes the credits already collected in this node, without consuming any extra cost.

Thus, based on the accounting methods, we prove that the cost for $N$ insertions is $O(N \cdot \log^2 N)$, which contains the cost of adjustment during those insertions, and the amortized cost for each insert operation is $O(\log^2 N)$ at most. □

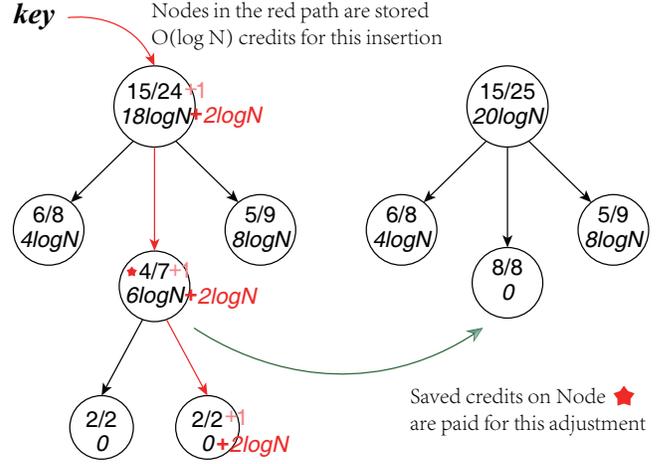

Figure 12: Amortized Analysis

# B MORE EXPERIMENTS

## B.1 Comparisons on Synthetic Datasets

Figure 13 displays the results of our method with different kernel functions compared with other baselines on synthetic datasets. We use different generating functions to construct synthetic datasets with 100M keys. These generating functions include Square (pow2), Cubic (pow3), Quartic (pow4), Logarithmic (log) and Exponential (exp) functions. Under these synthetic datasets, we compare the throughput of LIPP with other baselines, as well as LIPP equipped with specified kernel functions.

In Figure 13, LIPP equipped with the kernel function specifically chosen consistently performs better than the default one on different workloads. In some extreme cases, the throughput of LIPP with the specific kernel function is more than five times as that of the default setting, which is the significant performance improvement. On these synthetic datasets, LIPP with the default linear kernel function triggers more conflicts and adjustments especially on the dense and irregular area of the distribution. In this case, the number of tree levels also increases and the overall performance for both lookup and insert operations degrades. However, using a kernel function consistent with the data distribution can distribute the data more evenly, thereby reducing the element conflicts, the frequency of structure modifications and the number of levels in LIPP. Therefore, we provide the opportunity for users to inform LIPP the distribution of datasets by specifying the kernel function, so as to obtain more performance improvements.

Besides, the experimental results also indicate that our method (even only with the default kernel function) still outperforms the state-of-the-art methods on synthetic datasets with extremely non-linear and non-uniform distribution, which verifies the effectiveness of our method. In fact, even on wired and irregular datasets, our adjustment strategy can quickly balance the index tree and make the lookup and insert operations behave well in most cases. Besides, the adjustments have lower complexity and often happen on a relatively small collection of data, and thus do not incur too much overhead.

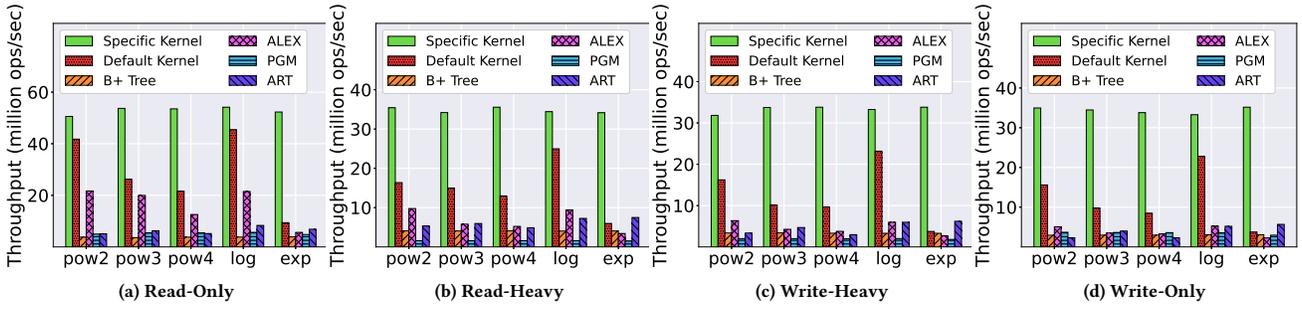

Figure 13: Comparisons on Synthetic Datasets

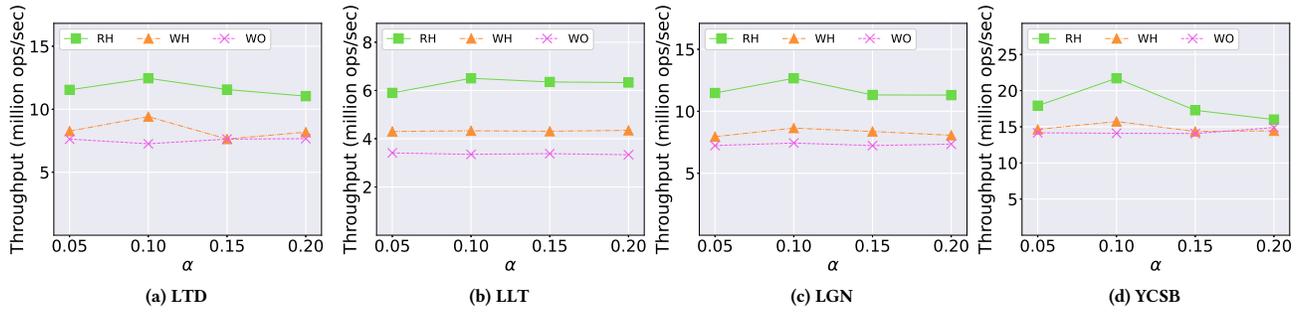

Figure 14: The Effect of $\alpha$

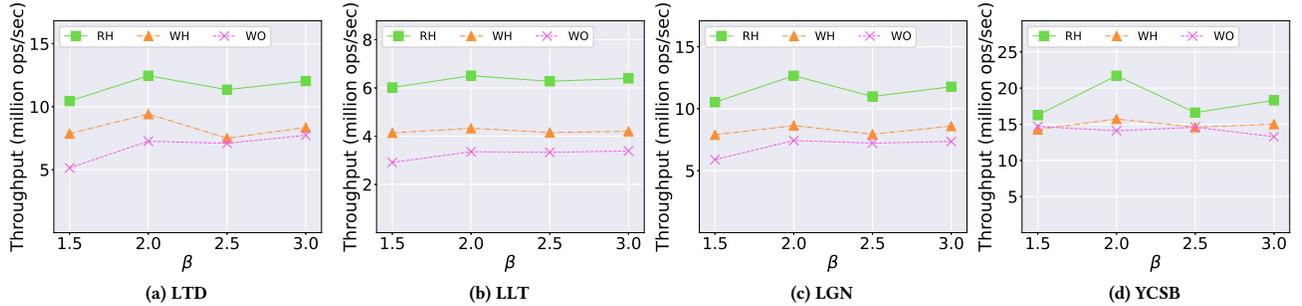

Figure 15: The Effect of $\beta$

Only on the datasets generated by the exponential functions does LIPP with the default kernel function behave worse than ART. But LIPP is still better than other baselines. In fact, the exponential generating functions make the first part of the generated dataset densely distributed, which causes repeated key collisions and adjustments. Thus, LIPP on datasets generated by the exponential function behaves relatively worse. Moreover, the datasets generated by the exponential function make ALEX frequently shift element to generate gaps for inserted elements and split nodes to adapt to the denser part of datasets.

### B.2 The Effect of Parameters

To choose the best values of parameters, we also give an analysis for parameters in LIPP, i.e., the threshold for conflict ratio $\alpha$ and the threshold for expansion ratio $\beta$ mentioned in Section 4.3.1. The results in Figure 14 and 15 show that the performance of LIPP using the default values for these parameters consistently is better than the performance of LIPP using other values even on different datasets and workloads, which proves the optimality of default values that LIPP chooses.

Though the default values of $\alpha$ and $\beta$ can bring some performance gains, we find that LIPP is not very sensitive to these parameters. In Figure 14, we find that there is at most 20% performance degradation for $\alpha = 0.2$ compared with the case when $\alpha = 0.1$. The similar trend of $\beta$ is shown in Figure 15 as well. Therefore, these situations indicate the stability of LIPP.

|      | LIPP | ALEX | B+Tree | ART  |
|------|------|------|--------|------|
| LTD  | 2.04 | 2.00 | 7.00   | 5.93 |
| LLT  | 2.99 | 2.23 | 7.00   | 6.20 |
| LGN  | 2.11 | 1.99 | 7.00   | 4.45 |
| YCSB | 1.63 | 1.00 | 7.00   | 4.46 |

Table 6: The Average Tree Height

## B.3 The Analysis for ALEX

We study the behaviour of ALEX to further demonstrate why ALEX behaves better than LIPP. In Table 4a, we show the average number of memory accesses used in the exponential search (Search) for ALEX on read-only workloads and in the elements shifting (Shift) to make gaps for ALEX on write-only workloads. The results show that ALEX costs about 30% memory accesses on the exponential search in leaf nodes for lookup operations. As for insert operations, ALEX costs more than 50% memory accesses in shifting elements and managing gapped array on most of datasets. However, those memory accesses for ALEX are not necessary for LIPP. In fact, LIPP saves the cost for those unnecessary memory accesses due to the precise position properties, thus having better performance.

(a) # Internal Memory Accesses

|      | Search | Shift |
|------|--------|-------|
| LTD  | 2.82   | 6.97  |
| LLT  | 2.15   | 11.95 |
| LGN  | 1.99   | 10.28 |
| YCSB | 3.30   | 3.66  |

(b) % Structure Modifications

|      | % Modification |
|------|----------------|
| LTD  | 12.7%          |
| LLT  | 14.7%          |
| LGN  | 13.8%          |
| YCSB | 14.0%          |

Table 5: The Analysis for ALEX

In Table 4b, we show the ratio of time spent on structure modifications to the overall running time for ALEX only on write-only workloads. The structure modifications for ALEX include four parts: Expand Scale, Expand Retrain, Split Sideways and Split Downwards. Detailed information can be found in paper [11]. Compared with Table 4, we find the ratios of time spent on the adjustments for LIPP and the structure modifications for ALEX are comparable, which means that the existing learned indexes still cost much time on the rebuilding or structure modifications. Besides, since the overall running time of ALEX is much larger than that of LIPP, the time spent on the adjustments for LIPP is less than the time spent on the structure modifications for ALEX on most datasets and workloads, even though the ratio for LIPP is larger. Thus, compared with ALEX, the cost of adjustments for LIPP is not the bottleneck and is acceptable for users.

## B.4 The Average Tree Height

Finally, Table 6 shows the tree height of different indexes on the write-only workloads, i.e., after inserting all elements from scratch. The tree height is averaged over keys. Since PGM uses logarithm methods to support insertions and has different trees, we simply ignore the tree height for PGM here. Learned Index does not support insert operations and cannot run on the write-only workloads. Thus, we also ignore the comparison with Learned Index.

The results of LIPP and ALEX show that datasets which are easier to model result in lower tree height. Besides, the tree height of LIPP is comparable with that of ALEX, which verifies that the tree height is not the essential factor for memory accesses and overall performance of LIPP and ALEX. Moreover, we find that the tree height of LIPP is far less than that of B+Tree and ART. Thus, when traversing the index trees, LIPP reach out the leafs more quickly than B+Tree and ART, and has fewer memory accesses and better overall performance. Also, B+Tree and ART need to search the keys in nodes of different layers, which incur more overheads.